\documentclass[12pt,fleqn]{article}
\usepackage{graphicx}
\usepackage{amsmath}

\def\ra{\rightarrow}

\usepackage[small]{caption}
\usepackage[usenames,dvipsnames]{color}
     \definecolor{hgreen}{rgb}{0,.3,0}
     \definecolor{hred}{rgb}{.3,0,0}
     \definecolor{hblue}{rgb}{0,0,.3}
     \definecolor{LightGray}{gray}{0.95}  

\usepackage[colorlinks=true,
            linkcolor=hblue,
            citecolor=hgreen,
            filecolor=hblue,
            urlcolor=hred]{hyperref}

\usepackage{relsize}
\def\babar{\mbox{\slshape B\kern-0.1em{\smaller A}\kern-0.1em
    B\kern-0.1em{\smaller A\kern-0.2em R}}}

\newcommand\pubnumber{}
\newcommand\pubdate{April, 2011}

\def\munich{
$^a$ Institute for Advanced Study, Technische Universit\"at M\"unchen, \\
Lichtenbergstra\ss{}e 2a, D-85748 Garching, Germany\\
\& Excellence Cluster Universe, Technische Universit\"at M\"unchen \\
Boltzmannstra\ss{}e 2, D-85748 Garching, Germany
}
\def\london{
$^b$ Blackett Laboratory, Imperial College London, \\
Prince Consort Rd, London SW7 2BW
}
\def\montreal{
$^c$ Institute of Particle Physics and \\
Department of Physics, McGill University \\
3600 rue University, Montr\'eal, Qu\'ebec
Canada H3A 2T8 
}

\long\def\symbolfootnote[#1]#2{\begingroup%
\def\thefootnote{\fnsymbol{footnote}}\footnote[#1]{#2}\endgroup} 

\def\mailg{\symbolfootnote[1]{mgorbahn@ph.tum.de}}
\def\mailp{\symbolfootnote[2]{mitesh.patel@imperial.ac.uk}}
\def\mailr{\symbolfootnote[3]{steven@physics.mcgill.ca }}

\def\Title#1{\begin{center} {\Large #1 } \end{center}}
\def\Author#1{\begin{center}{ \sc #1} \end{center}}
\def\Address#1{\begin{center}{ \it #1} \end{center}}

\newcommand\pubblock{\rightline{\begin{tabular}{l} \pubnumber\\
         \pubdate  \end{tabular}}}
\newenvironment{Abstract}{\begin{quotation}  }{\end{quotation}}
\newenvironment{Presented}{\begin{quotation} \begin{center} 
             PRESENTED AT\end{center}\bigskip 
      \begin{center}\begin{large}}{\end{large}\end{center} \end{quotation}}

\begin{document}
\begin{titlepage}
\pubblock

\vfill
\Title{Summary of the CKM 2010 Working Group on Rare Decays}
\vfill
\Author{
  Martin Gorbahn$^{a}$\mailg, 
  Mitesh Patel$^{b}$\mailp and 
  Steven Robertson$^{c}$\mailr}
\Address{\munich}
\Address{\london}
\Address{\montreal}
\vfill

\begin{Abstract}
  Rare decays were essential in the discovery of the CKM mechanism of
  flavour and CP violation and are highly sensitive probes of physics
  beyond the Standard Model. In this summary the current status and
  future prospects of experimental measurements and the Standard Model
  theory predictions of various rare $B$, $D$ and $K$ decay
  observables are discussed. The specific new physics sensitivities of
  each mode are also briefly reviewed.
\end{Abstract}

\vfill

\begin{Presented}
CKM 2010\\
the $6^{th}$ International Workshop on the CKM Unitarity Triangle\\
University of Warwick, UK, 6-10 September 2010
\end{Presented}
\vfill
\end{titlepage}
\def\thefootnote{\fnsymbol{footnote}}
\setcounter{footnote}{0}

\section{Introduction}

Historically, the study of rare decays was instrumental in the
discovery of the Cabibbo Kobayashi Maskawa (CKM) picture of quark
mixing and the prediction of the existence and properties of
heavy fermions \cite{Glashow:1970gm,Kobayashi:1973fv}. Recent
experimental progress has confirmed the CKM mechanism as the dominant
source for flavour and CP violation.
In the coming years, the focus will again shift to the discovery of new
heavy particles. Rare decays can provide an essential tool in
measuring the coupling constants and mixing angles of any new, heavy
degrees of freedom, which in turn will reveal the symmetries of the
underlying Lagrangian.
Observables which
can be predicted with high precision, are experimentally accessible,
and have an enhanced sensitivity to new physics will play a central role.
In the following, the findings of the CKM working group are summarized
and the above criteria are discussed mode-by-mode.

In this working group flavour changing neutral current (FCNC)
radiative and leptonic decays of $B$, $D$ and $K$ mesons were
discussed. At leading loop order all new physics (NP) models with
heavy particles discussed here can be described at the mass scale of
the decaying meson by the effective Hamiltonian
\begin{equation}
  \mathcal H_\text{eff}^{d_i\to d_j} = 
  4 \frac{G_F}{\sqrt{2}}V_{ti}V_{tj}^*
  \sum_k(C_k^{ij}Q_k^{ij}+C_k^{\prime ij}Q_k^{\prime ij})
  \,,
\label{eq:heff}
\end{equation}
where $V_{ij}$ is the CKM matrix and the operators relevant at tree
level are
\begin{align}
Q_{7}^{(\prime)ij} &= 
m_b
(\bar d_j \sigma_{\mu \nu} P_{R(L)} d_i) F^{\mu \nu} ,&
Q_{8}^{(\prime)ij} &= 
m_b (\bar d_j \sigma_{\mu \nu} T^a P_{R(L)} d_i) G^{\mu \nu \, a} ,&
\nonumber\\
Q_{9}^{(\prime)ij} &= 
(\bar d_j \gamma_{\mu} P_{L(R)} d_i)(\bar{\ell} \gamma^\mu \ell) ,&
Q_{10}^{(\prime)ij} &= 
(\bar d_j  \gamma_{\mu} P_{L(R)} d_i)(  \bar{\ell} \gamma^\mu \gamma_5 \ell) ,&
\nonumber\\
Q_{S}^{(\prime)ij} &= 
m_b (\bar d_j P_{R(L)} d_i)(  \bar{\ell} \ell) ,&
Q_{P}^{(\prime)ij} &= 
m_b (\bar d_j P_{R(L)} d_i)(  \bar{\ell} \gamma_5 \ell) ,&
\nonumber\\
Q_{L(R)}^{ij} &= 
(\bar d_j \gamma_\mu P_{L(R)} d_i)(  \bar{\nu} \gamma^\mu P_L \nu) \, .&
\label{eq:ops}
\end{align}
$P_{L(R)}$ denotes the projector of the left-handed (right-handed)
field for the (primed) operator. In the Standard Model (SM) only the
Wilson coefficients $C_k^{ij}$ of the unprimed operators are
generated.  The small size of the light-quark Yukawa couplings
suppresses the Wilson coefficients $C_{S,P}$ of the scalar operators
even further and renders their contribution negligible.

In extensions of the SM these suppression mechanisms need not be
present. A strong SM suppression increases the new physics sensitivity
and there are several suppression mechanisms at work: loop suppression
is present in all FCNC processes and new heavy degrees of freedom can
contribute at the same level as the SM. Helicity suppressed modes are
particularly sensitive to flavour changing scalar interactions and CKM
suppressed modes to new sources of flavour violation.

In order to be able to exploit this new-physics sensitivity, a precise
knowledge of the SM background is needed. To calculate this
background, matrix elements of four-quark operators must be calculated
at higher loop order. In this case the current-current operators
$Q_{1,q} = (\bar d_j \gamma_\mu P_L q) (\bar q \gamma_\mu P_L d_i)$
and $Q_{2,q} = (\bar d_j \gamma_\mu T^a P_L q) (\bar q \gamma_\mu T^a
P_L d_i)$ play the most important role (for decays of down-type quarks
$q$ represents the up and charm quarks, for charm decays it stands for
down and strange quarks -- $T$ denotes the Gell-Mann matrices). Their
Wilson coefficients are tightly constrained and usually insensitive to
physics beyond the SM. However, the calculation of their matrix
elements often results in a sizeable uncertainty in the estimation of
the SM background.  Contributions of higher-order matrix elements can
often be absorbed in a redefinition of $C_i$ in terms of effective
Wilson coefficients $C_i^{\textrm{eff}}$.

\section{$b \to s \ell^+ \ell^-$ and $b \to s \gamma$ decays }

The decays $b \ra s \ell^+ \ell^-$, where $\ell^+ \ell^-$ is an $e^+
e^-, \mu^+ \mu^-$ or $\tau^+ \tau^-$ pair, are FCNC processes which
occur via a $b\to s$ transition through a loop diagram.  In the SM the
Wilson coefficient $C_7$ plays the dominant role for $b \to s (d)
\gamma$ decays. In the case of $b \to s (d) \ell^+ \ell^-$ decays the
coefficients $C_9$ and $C_{10}$ also contribute signifcantly.  In
extensions of the SM the chirality flipped operators $Q'_7$, $Q'_9$
and $Q'_{10}$, as well as the scalar operators can also contribute. A
range of observables sensitive to specific combinations of Wilson
coefficients of these operators exist. Apart from the integrated
branching ratio, charged lepton forward-backward asymmetries (${\cal
  A}_{FB}$) as well as isospin and CP asymmetries can be defined for
both the exclusive and inclusive decay modes. The dependence of the $b
\to s (d) \ell^+ \ell^-$ decay rate on the invariant mass $q^2$ of the
lepton pair is described by the differential decay rate. The $q^2$
region dominated by charm-quark resonances is usually excluded and the
total branching ratios are given for the regions above or below the
resonances.  The exclusive decays offer even more observables: in
particular the angular analysis of the $B \to K^* \ell^+ \ell^-$
decays, where the $K^{*}$ polarisation is measured using the $K^{*}
\to K^- \pi^+$ decay, leads to 24 observables for each leptonic mode
per $q^2$-bin.

Inclusive $B$ decays can be reliably approximated by perturbative
calculations. The large bottom-quark mass $m_b$ allows the decay rates
to be expanded in terms of the partonic decay rate for $b \to s
\gamma$, i.e. $\Gamma(B \to X_s \gamma) = \Gamma(b \to s \gamma) +
\mathcal{O}(\Lambda_\textrm{QCD}/m_b)$. A similar expansion holds for
$b \to s \ell^+ \ell^-$. For both partonic decay rates the NNLO
calculations are almost complete \cite{Misiak:2006zs,Bobeth:2003at} --
see Ref.~\cite{Paz:2010wu} for a summary of recent developments. The
non-perturbative uncertainty of $\Gamma(B \to X_s \gamma)$ 
was discussed in the talk by Paz \cite{Paz:2010wu}: a study of the
$\mathcal{O}(\Lambda_\textrm{QCD}/m_b)$ uncertainty, including the
$Q_7 - Q_8$ contribution, results in an intrinsic 5\% non-perturbative
uncertainty for the radiative mode. The CP asymmetry in the $b \to s
\gamma$ decay modes is small in the SM and a large
deviation from zero would be a signal of new physics. The
non-perturbative uncertainty of the CP asymmetry was recently studied
in \cite{Benzke:2010tq}, where the non-perturbative matrix element of
$Q_1 - Q_7$ lead to 2\% long-distance pollution, which is larger than
previously assumed.

Exclusive decays provide an even richer phenomenology, in particular
for the many physical observables associated with the $B \to K^*
\ell^+ \ell^-$ decay. Yet, a direct calculation of the weak decay
matrix elements is mandatory for a precise theory prediction. Here
factorisation (and soft collinear effective theory) provides a
framework to disentangle perturbatively calculable hard-scattering
kernels \cite{Beneke:2004dp} from form factors and light-cone
distribution amplitudes in a power expansion in
$\mathcal{O}(\Lambda_\textrm{QCD}/m_b)$.

The current uncertainty for the form factors of $B$ decays into
pseudo-scalar or vector particles, is 12 --
15\%\cite{Khodjamirian:2011gy}. Form factors in the region of small
and intermediate $q^2$ are calculated from QCD light-cone sum rules --
see Ref.~\cite{Khodjamirian:2011gy} for a summary of recent
activities. In the high $q^2$ region improvements can be made using
lattice QCD. The recent progress in the calculation of the $B \to
K^{(*)}$ form factors with a moving-non-relativistic QCD action is 
presented in Ref.~\cite{Liu:2011ra}.

\begin{figure}[t]
\includegraphics[scale=0.7]{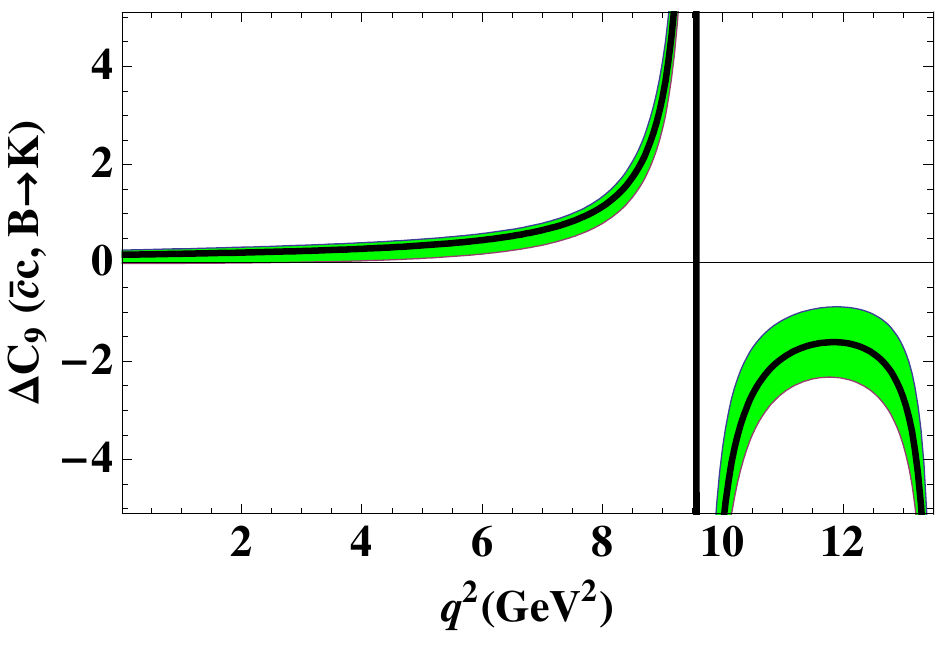}
\hspace{2mm}
\includegraphics[scale=0.7]{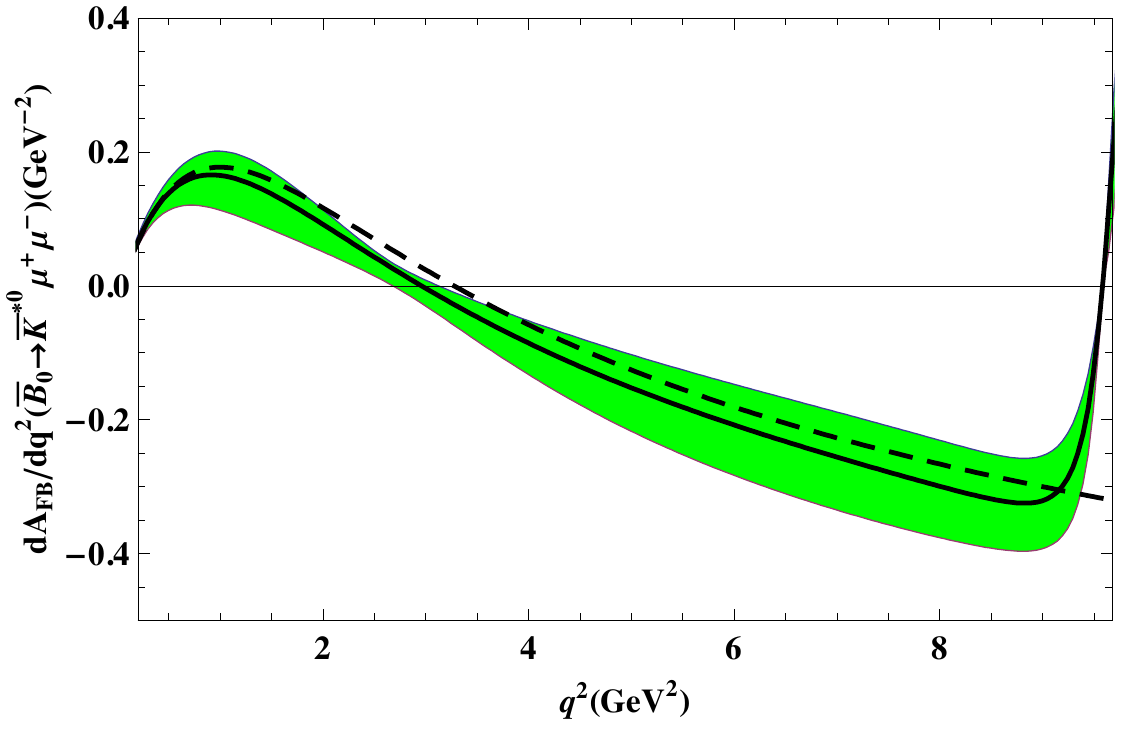}
  \caption{\label{khod1} (Left) The charm loop contribution to the
    Wilson coefficient $C_9$ for $B \ra K^* \ell^+ \ell^-$ decays. The
    central values are denoted by the dashed line and the shaded area
    indicates the estimated uncertainties. (Right) The
    forward-backward asymmetry ${\cal A}_{FB}$ with (solid) and
    without (dashed) the effect of the charm-loop. Taken from
    Ref.~\cite{Khodjamirian:2010vf}.}
\end{figure}

Beyond the leading-order factorisable contributions and hard gluon
exchanges there also exist non-factorisable contributions to $B \to
K^{(*)} \ell^+ \ell^-$. A soft gluon which couples to the lepton pair
via a virtual charm loop, which is induced by a $Q_{1/2}$ operator
insertion, cannot be expressed in terms of a local operator.  The
influence of the resulting non-local operator
\cite{Khodjamirian:2011gy,Khodjamirian:2010vf} is illustrated in
Fig.~\ref{khod1}, and can be absorbed in a correction to the
effective coefficient of the $O_{9}$ operator. The impact on the
${\cal A}_{FB}$ observable is also shown.

\begin{figure}[t]
  \includegraphics[height=.33\textwidth,width=.5\textwidth]{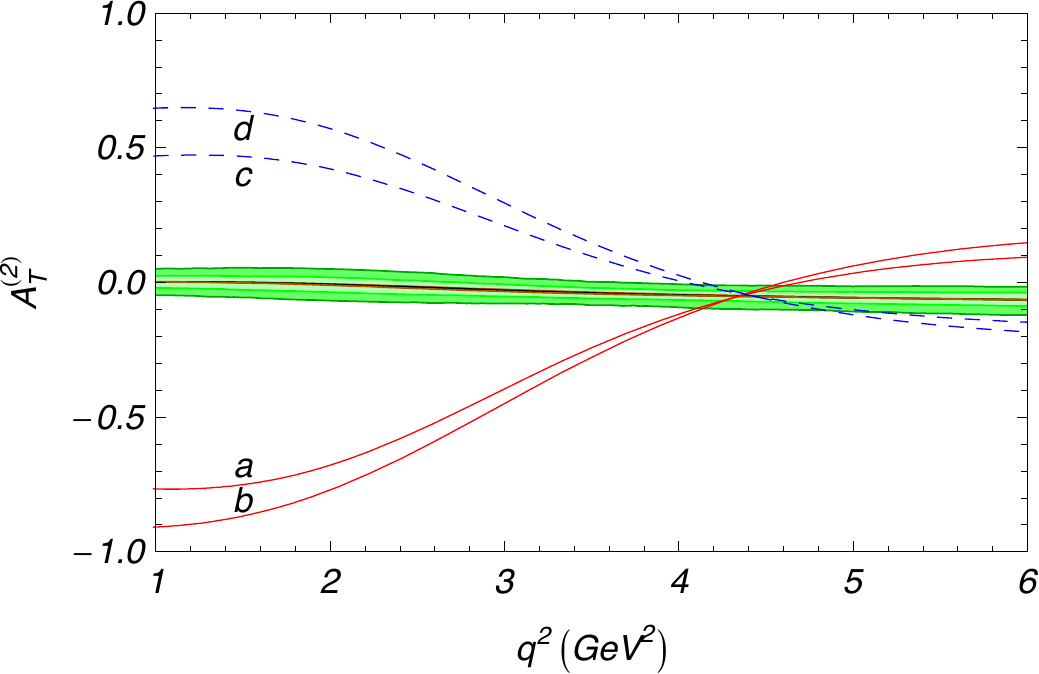}
  \includegraphics[height=.33\textwidth,width=.5\textwidth]{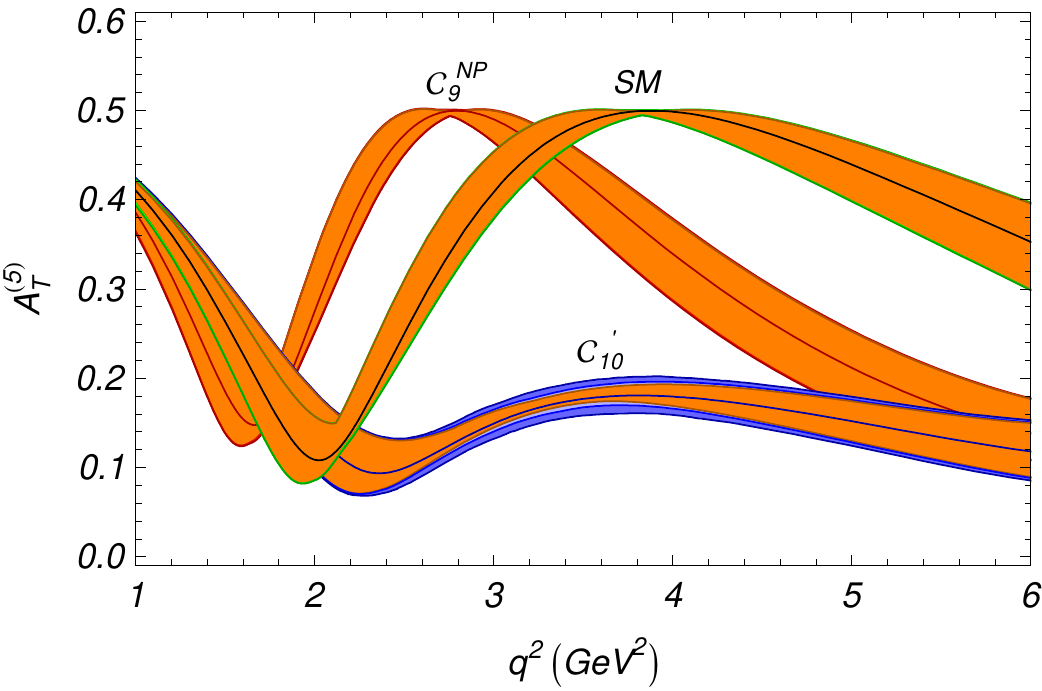}
  \caption{(Left) $A_T^2$ in SM (green band) with four NP benchmarks
    (\cite{Egede:2010zc}). (Right) $A_T^5$ in the SM and for different
    values of $C_9^{eff}$ and $C_{10}^\prime$. For more details see
    Ref.~\cite{Egede:2010zc}.}
\label{fig:magnet}
\end{figure}

The current uncertainty in the form factors is the limiting
uncertainty in the theory prediction of the radiative and electroweak
rare decays.  Given the difficulties of making precise theoretical
predictions for the exclusive decay modes, attention has therefore
focused on observables with a reduced dependence on the form-factors.
The zero-crossing point of ${\cal A}_{FB}$ is the most well known
example of a quantity which is almost free of such hadronic
uncertainties. However, recent theoretical work has identified a
number of additional observables. Quantities such as $A_T^2$ and
$A_T^5$, which are defined and discussed in detail in
Ref.~\cite{Egede:2010zc}, have precise theoretical predictions
(Fig.~\ref{fig:magnet}). The $A_T^2$ observable contains almost all
the physical information of ${\cal A}_{FB}$ but is less sensitive to
QCD uncertainties and exhibits a much larger sensitivity to
right-handed currents.
 
The \babar, Belle and CDF experiments have all measured exclusive $b
\ra s \ell^+ \ell^-$ decays~\cite{babar, belle, cdf, Eigen:2011rq},
where the final lepton pair state comprises electron and muons for the
B-factories and only muons at CDF.  Events are selected from data
samples with integrated luminosities of $349~ \rm fb^{-1}$, $605~ \rm
fb^{-1}$ and $4.4~ \rm fb^{-1}$ respectively, corresponding to 384
million $B \bar B$ events, 656 million $B \bar B$ events and $2 \times
10^{10} ~b \bar b$ events.  The measured total branching ratio is in
good agreement with theoretical predictions, as is the direct CP
asymmetry -- see Ref.~\cite{Eigen:2011rq} for a summary table of
recent results.  Further measurements have therefore focussed on
additional observables involving the angular
distributions~\cite{Burdman:1998mk,Beneke:2001at,Alok:2009tz} or
isospin asymmetries~\cite{Beneke:2004dp,Feldmann:2002iw}.  The \babar,
Belle and CDF experiments have also analyzed the one-dimensional
angular distributions which involve the observables ${\cal F}_L $, the
fraction of $K^*$ longitudinal polarization, and ${\cal A}_{FB}$, the
lepton forward-backward asymmetry.

\begin{figure}[t]
\centering
 \hskip -0.3cm 
 \includegraphics[height=1.75in]{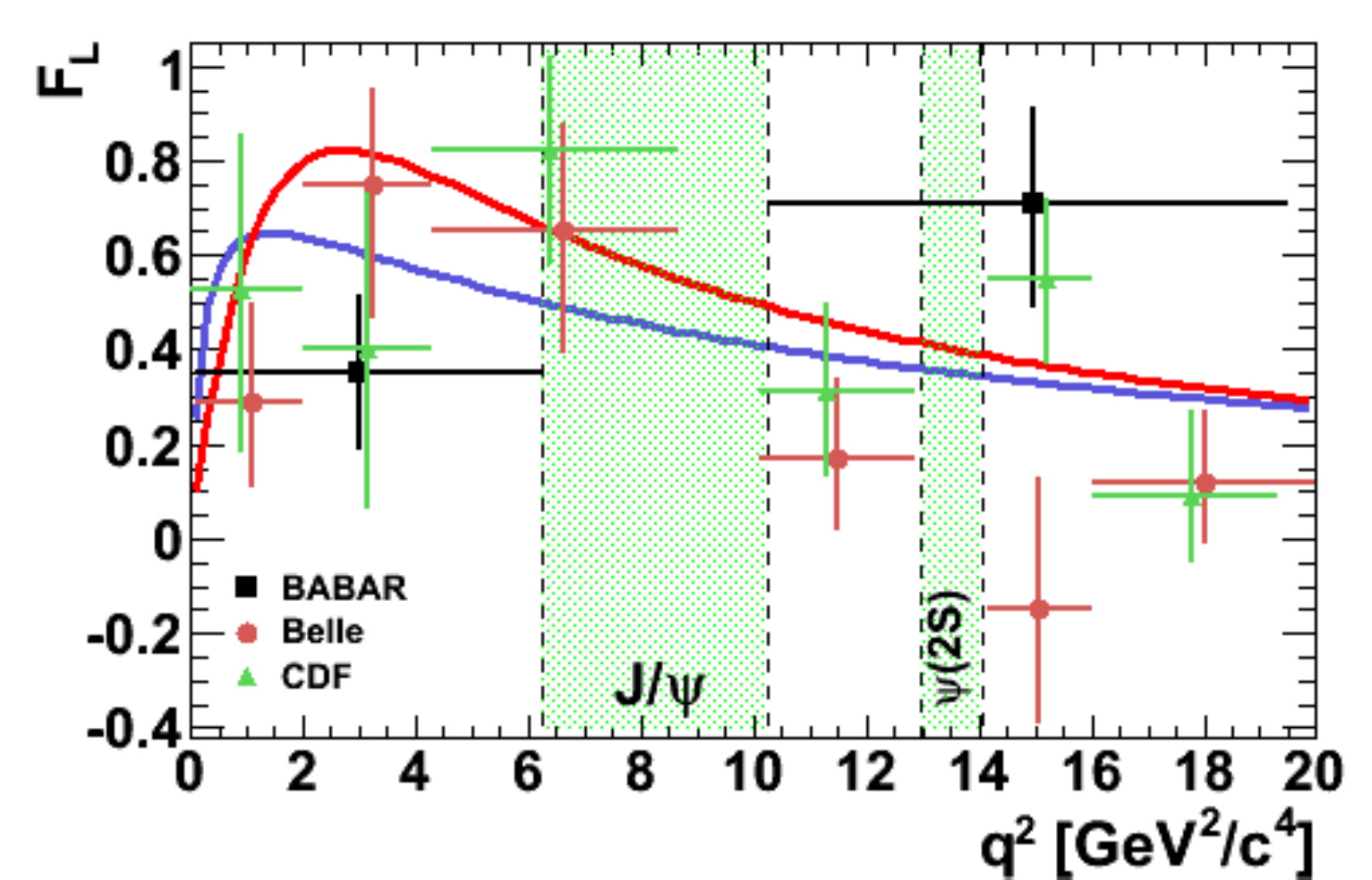} 
\hskip -0.3cm
 \includegraphics[height=1.75in]{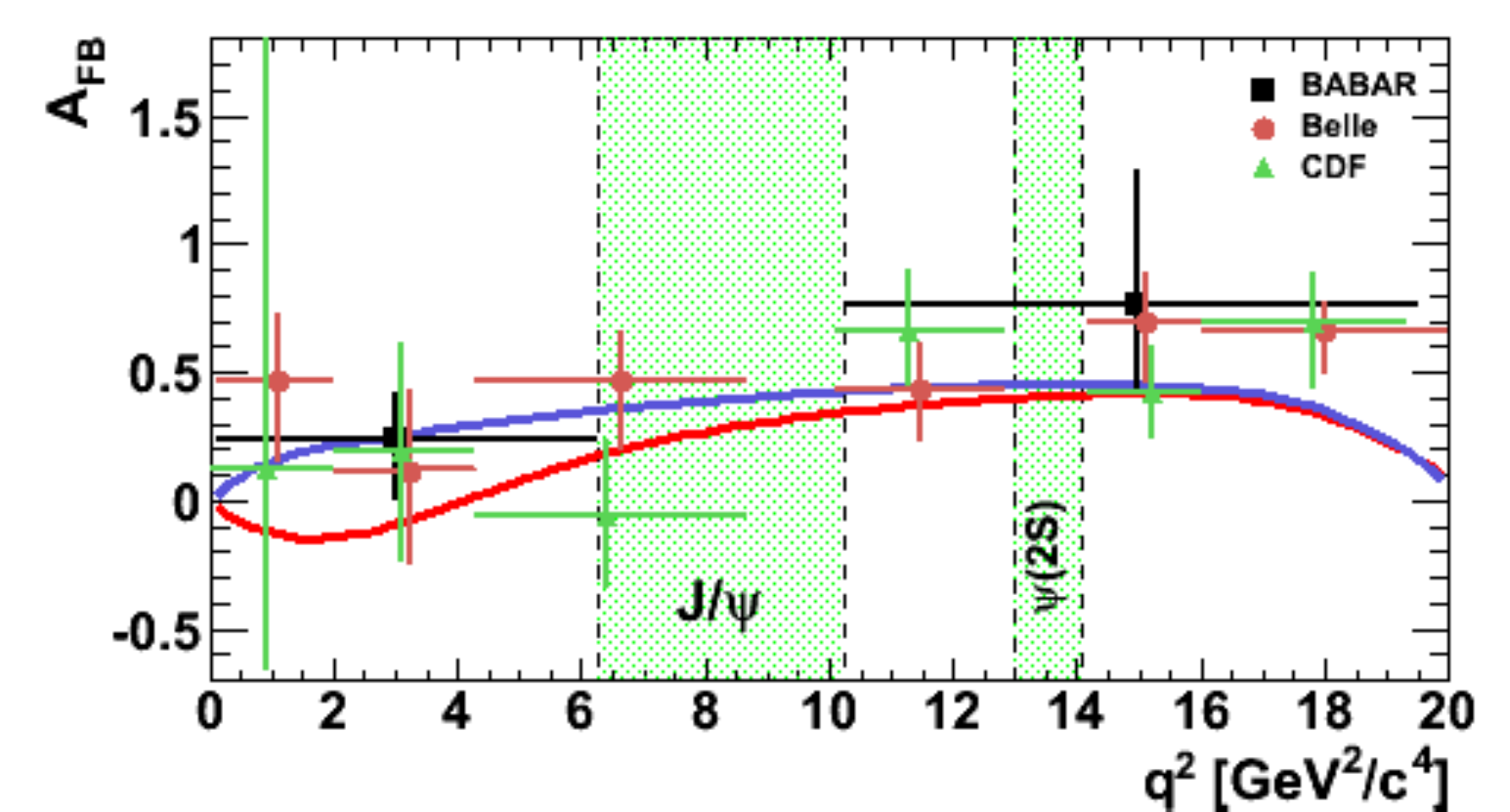}
\caption{(Left) Measurements of ${\cal F}_L$ and (Right) ${\cal
    A}_{FB}$ in $B \ra K^{(*)} \ell^+ \ell^-$ decays by the \babar\
  (black squares), Belle (brown dots) and CDF (green triangles)
  experiments. The SM prediction (flipped-sign $C^{eff}_7$ model) is
  shown by the upper red (lower blue) curve for ${\cal F}_L$ and the
  lower red (upper blue) curve for ${\cal A}_{FB}$. 
The green shaded regions show the $J/\Psi$ veto used by Babar and the $\Psi(2S)$ veto used by Belle and CDF.
}
\label{fig:afb1}
\end{figure}

Fig.~\ref{fig:afb1} shows the \babar~\cite{babar}, Belle~\cite{belle},
and CDF~\cite{cdf} results for ${\cal F}_L$ (left) and ${\cal A}_{FB}$
(right) as a function of the di-muon invariant mass squared,
$q^2$. The figure also shows the SM prediction (lower red
curve)~\cite{buchalla} and that of a model in which the sign of
$C^{eff}_7$ is flipped (upper blue curve)~\cite{hou, kruger2}. The
measurements from all three experiments are in good agreement. While
the measurements are in better agreement with the flipped-sign
$C^{eff}_7$ model, they are consistent with the SM prediction. For $B
\ra K \ell^+ \ell^-$, ${\cal A}_{FB}$ is consistent with zero, as
expected in the SM.

\begin{figure}[t]
\centering
 \includegraphics[height=1.75in]{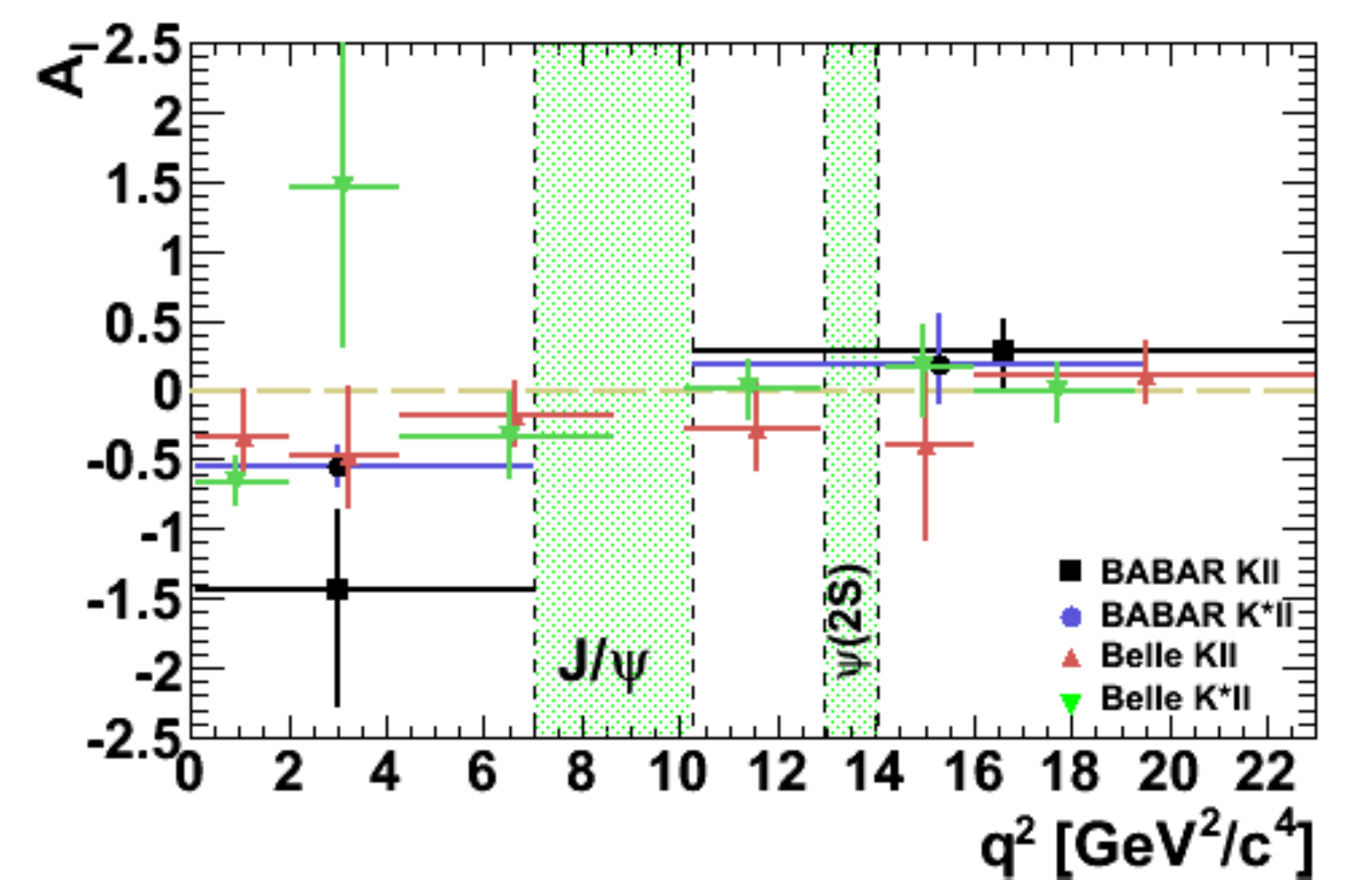}
\caption{Isospin asymmetry measurements for $B \ra K^{(*)} \ell^+ \ell^-$  versus $q^2$ from \babar\ (black squares, blue dots) and Belle (red triangles, green triangles).}
\label{fig:bf}
\end{figure}

The isospin asymmetry, ${\cal A}_I$,
\begin{equation} {\cal A}_I (q^2)= \frac{d{\cal B}(B^0 \ra K^{(*)0}
    \ell^+ \ell^-)/dq^2 -(\tau_{B^0}/\tau_{B^+}) d {\cal B}(B^+ \ra
    K^{(*)+} \ell^+ \ell^-)/dq^2} {d{\cal B}(B^0 \ra K^{(*)0} \ell^+
    \ell^-)/dq^2 +(\tau_{B^0}/\tau_{B^+}) d {\cal B}(B^+ \ra K^{(*)+}
    \ell^+ \ell^-)/dq^2},
\end{equation}
is expected to be small in the SM. In particular, in the region $q^2
=2.7-6~\rm GeV^2/c^4$, the expectation for $d {\cal A}_I (q^2)/dq^2$
is $-0.01$ in the case of $B \to K^* \ell^+
\ell^-$~\cite{Feldmann:2002iw}.  Fig.~\ref{fig:bf} shows the \babar\
and Belle ${\cal A}_I$ measurements as a function of $q^2$. Both the
$q^2$-integrated isospin asymmetry, and the ${\cal A}_I$ value for
$q^2$ larger than the $J/\psi$ invariant mass-squared, are consistent
with the SM prediction. However, below the $J/\psi$ mass-squared,
\babar\ observes a negative ${\cal A}_I$ which is $ 3.9 \sigma$ from
the SM prediction of ${\cal A}_I=0$. For low $q^2$, the Belle results
are consistent with both the \babar\ measurements and with the SM
expectation.

\begin{figure}[t]
\centering
   \includegraphics[height=.33\textwidth,width=.33\textwidth]{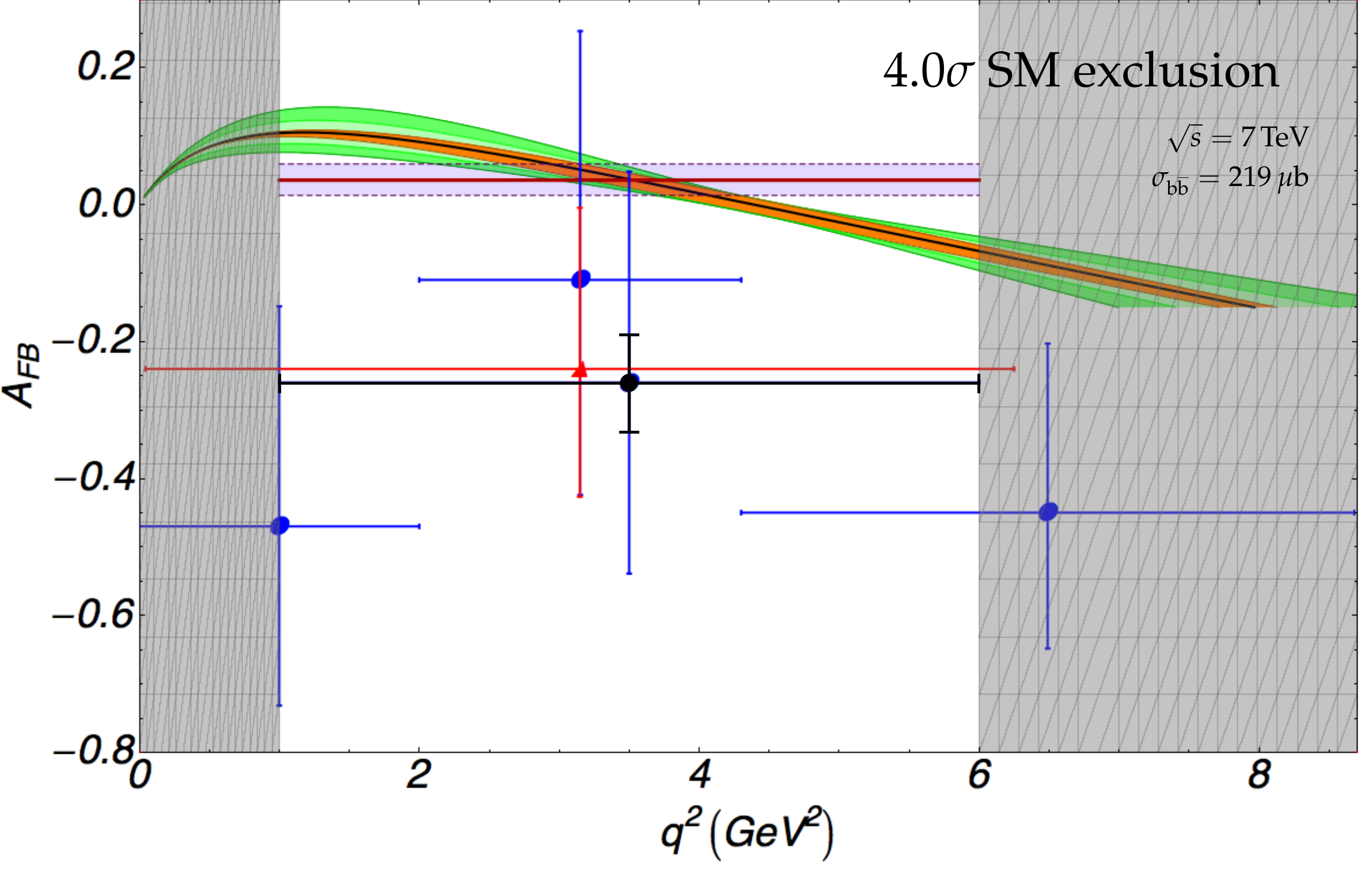}
  \caption{The precision on ${\cal A}_{FB}$ expected from the LHCb
    experiment with $1\rm fb^{-1}$ integrated luminosity. The present
    central value given by the Belle experiment is shown with a purely
    statistical uncertainty. The shaded bands show the theory
    expectation in the SM with uncertainties (orange), additional
    $\Lambda/m_B$ corrections at the 10\% level (green) and the
    average over the $1<q^2<6\,$GeV$^2$ region (pink). Taken from
    Ref.~\cite{Skottowe:2010bs}.}
\label{fig:afb}
\end{figure}

The $B$-factory measurements of ${\cal A}_{FB}$ will be improved by
the LHCb experiment. Fig.~\ref{fig:afb} shows the precision on ${\cal
  A}_{FB}$ expected with $1\rm ~fb^{-1}$ of integrated
luminosity~\cite{Soomro:2011vn}. This dataset will allow the
zero-crossing point of ${\cal A}_{FB}$ to be determined with 13\%
precision~\cite{roadmap}. Observables such as $A_T^2$ can be extracted
from the projection of the distribution of a single angle. With
sufficient experimental data it will be possible to make a full
angular fit and improve the precision on this and other
observables. This will also allow those observables requiring
information from multiple angular distributions to be determined.

Experimental observables for $b \to s \gamma$ and $b \to d \gamma$
include the inclusive and exclusive branching fractions, direct CP
asymmetry
\begin{equation}
A_{CP} = \frac{\Gamma (b \to s \gamma) - \Gamma (\bar{b} \to \bar{s} \gamma)}{\Gamma (b \to s \gamma) + \Gamma (\bar{b} \to \bar{s} \gamma)}
\end{equation}
and the shape of the photon energy spectrum.  The photon energy
spectrum is not sensitive to new physics, but can be used to determine
Heavy Quark Expansion parameters related to the $ b$ quark mass and
momentum within the $B$ meson. These are used in the extraction of
$|V_{ub}|$ and $|V_{cb}|$ from semileptonic $B$ decays.  The inclusive
$b \to s \gamma$ branching fraction has been computed to NNLO to a
precision of about $7\%$~\cite{Misiak:2006zs} for $E_\gamma >
1.6$~GeV, comparable in precision to current experimental
determinations.

Inclusive measurements of the $b \to s \gamma$ branching fraction are
based on three distinct techniques: (1) summing many exclusive decay
modes, (2) fully inclusive measurements utilising kinematic
information to suppress backgrounds, or (3) exclusive reconstruction
of the second $B$ meson in the event.  Each method has its own
strengths and weaknesses and in practice all three are competitive
with current $B$-factory data statistics.  In general, none of these
methods explicitly reject $b \to d \gamma$, hence this contribution is
usually implicitly included in the branching fraction and $A_{CP}$
definitions.  The most precise branching fraction measurement comes
from Belle~\cite{belle_incl_bsgamma} based on $605$~fb$^{-1}$, with
$B(B\to X_s \gamma) =(3.45 \pm 0.15 \pm 0.40) \times 10^{-4}$ for
$E_\gamma > 1.7$~GeV.  The combined world average branching fraction
is consistent with the NNLO theory prediction at roughly the $1\sigma$
level.  The \babar\ experiment has recently reported a preliminary
determination of $A_{CP}$ based on a fully inclusive
method~\cite{wenfang} and utilizing a data sample of $347$~fb$^{-1}$.
To reduce background contamination, the search is restricted to a
limited region of the photon energy spectrum, $2.1 < E_{\gamma}
<2.8$~GeV, resulting in $A_{CP} = 0.056 \pm 0.060 \pm 0.018$.  This
measurement is consistent with no CP asymmetry and is significantly
more precise than previous measurements.

Exclusive and sum-of-exclusive determinations of the $b \to d \gamma$
branching fraction~\cite{belle_excl_bdgamma,
  babar_excl_bdgamma,babar_debbie} can be compared with the
corresponding $b \to s \gamma$ measurements to extract
$|V_{td}|/|V_{ts}|$~\cite{ball_bdgamma}.  This can be compared with
the more precise measurement of this quantity from $B_s$ and $B_d$
mixing.

Future $B$-factory facilities offer scope for substantial improvement
in inclusive (and semi-inclusive) $B \to X_{s,d} \gamma$ measurements
as well as the potential to perform fully inclusive studies of $B \to
X_s \ell^+ \ell^-$ and possibly $B \to X_d \ell^+ \ell^-$ .  Fully
inclusive $B \to X_{s,d} \gamma$ measurements are typically limited by
the enormous background from continuum events and from photons
originating from $\pi^0$ and $\eta$ mesons produced in $B$ decays.
Semi-inclusive methods based on summing a large number of exclusive
modes are able to more strongly suppress these backgrounds but suffer
from large uncertainties related to the normalization of the
individual modes and to the modes which are not reconstructed.  By
requiring exclusive hadronic ``tag'' reconstruction of the $B$ meson
accompanying the $B \to X_{s,d} \gamma$ signal candidate, not only is
the continuum background strongly suppressed, but the kinematics of
the signal $B$ are fully constrained, resulting in improvements in
resolution of the photon energy spectrum and hadronic $X_{s,d}$
system, as well as a strengthening of the $\pi^0$ and $\eta$ veto. In
principle this method permits a truly inclusive measurement over the
entire $X_{s,d}$ mass range, however current $B$-factory measurements
using this method are statistically limited to the extent that they
cannot compete with untagged measurements.  With the addition of two
orders of magnitude of additional data from the new super
$B$-factories, this method will have statistical sensitivity
comparable to current untagged analyses from \babar\ and Belle, but
with substantially reduced backgrounds and systematic uncertainties.
In the case of $B \to X_{s,d} \ell^+ \ell^-$, the hadronic tag
reconstruction would permit studies based only on the $\ell^+ \ell^-$
kinematics and hence accessing the entire $X_{s}$ spectrum, however
these measurements are not competitive using current $B$-factory data
samples. With super $B$-factory data samples of order $100$~ab$^{-1}$,
these measurements may prove complementary to LHCb exclusive
determinations.

\section{$B \to K^{(*)}\nu \bar{\nu}$}

The rare decays $b \to s \nu \bar{\nu}$, like their charged lepton
counterparts, provide an interesting potential window into new physics
in FCNC processes.  In contrast to $b \to s \ell^+ \ell^-$, there is
no photon penguin contribution and only the $Z$-penguin needs to be
considered.  Hence, in the SM, the effective Hamiltonian contains only
a single dimension-six operator and single complex Wilson coefficient
$C_L$.  Potential new physics contributions arise in many models
including the Minimal SuperSymmetric Model (MSSM), little Higgs and
extra dimensions models, and can result in a non-zero $C_R$ and non-SM
value of $C_L$.  These can be constrained by measurements of a
combination of inclusive and exclusive $b \to s \nu \bar{\nu}$
modes~\cite{Kamenik:2010na}.  In addition, since the neutrinos are not
observed, the experimental signature of $B \to K^{(*)} + E_{\rm miss}$
does not distinguish between the $\nu \bar{\nu}$ mode or other
unobserved final state particles~\cite{Smith:2010st}.  This can lead
to apparent deviations from the SM branching fraction or modifications
to the kinematic spectra of the $K^{(*)}$.

Although the inclusive $B \to X_s \nu \bar{\nu}$ processes are
theoretically very clean, in practice only the exclusive modes $B \to
K^{(*)}\nu \bar{\nu}$ are experimentally accessible at $B$-factories,
due to the very limited kinematic information available to constrain
these decays.  The experimental challenge is to ascertain that the
only visible decay daughter of the $B$ is a charged or neutral
$K^{(*)}$.  At $B$-factories, this is accomplished using exclusive tag
reconstruction of the other $B$ meson in the $\Upsilon(4S)\to
B\bar{B}$ event, in one of a large number of either fully-hadronic or
semileptonic decay modes.  Stringent kinematic constraints on the
reconstructed $B$ ensure the high purity of this reconstructed sample,
substantially suppressing $q \bar{q}$ continuum backgrounds and
mis-reconstructed $B\bar{B}$ decays.  Once a clean tag $B$ sample is
obtained, the remaining detected particles are hypothesised to
originate from a $B\to K^{(*)}\nu \bar{\nu}$ signal candidate.  Signal
events typically contain only the $K^{(*)}$, plus a small number of
low-energy calorimeter energy deposits originating from beam-related
backgrounds, hadronic shower fragments (``split-offs'') and similar
debris.  In the $K^{\pm}$ and $K_s^0$ modes the Kaon momentum
$p^*_{K^{(*)}}$ is the only physical observable, while in the $K^*$
modes the longitudinal or transverse polarization fraction, accessible
via the angular distributions of the $K^*$ decay products, is also
available.  In all modes significant backgrounds from generic $b \to
c$ processes contribute at low $p^*_{K^{(*)}}$, so current
experimental searches are mainly sensitivity to the high
$p^*_{K^{(*)}}$ region and in most cases impose a kinematic
requirement that $p^*_{K^{(*)}} > 1.5$~GeV$/c$.  This introduces a
modest form factor dependence on experimental limits, but can also
significantly impact sensitivity to specific new physics models which
predict a softer $p^*_{K^{(*)}}$ spectrum, for example if the missing
energy originates from a pair of undetected massive scalar particles.
Experimental strategies therefore necessitate a choice between the
need to suppress background and the desire to retain the largest
possible portion of the $p^*_{K^{(*)}}$ spectrum.  A recent \babar\
analysis~\cite{babar_SL_Knunu}, utilising a Bagged Decision Tree
multivariate approach, attempted to address this issue by reporting
separately a partial branching fraction limit in the
$p^*_{K^{(*)}}>1.5$~GeV$/c$ region and the low $p^*_{K^{(*)}}$ region.
However, due to the resulting strong dependence of the signal
efficiency on $p^*_{K^{(*)}}$, interpretation of this result in a new
physics context remains problematic.

\babar\ and Belle have published limits on the exclusive modes $B^+
\to K^+ \nu \bar{\nu}$, $B^0 \to K_s^0 \nu \bar{\nu}$, $B^+ \to K^{*+}
\nu \bar{\nu}$ and $B^0 \to K^{*0} \nu \bar{\nu}$, based on a
combination of the hadronic and semileptonic tag reconstruction
techniques~\cite{babar_SL_Knunu, exp_knunu}.  These searches currently
limit the branching fractions to a few times the SM rates and are
essentially statistically limited by the low tag reconstruction
efficiency.  Further improvements in these modes will require
additional statistics, either due to substantial improvements in the
tag reconstruction method at the current generation of $B$-factories,
or due to new data recorded at future Super $B$-factory facilities in
Italy and Japan.

\section{$B_q \to \ell^+ \ell^-$ and $D \to \ell^+ \ell^-$}

The rare decay $B_q \to \ell^+ \ell^-$ results in a final-state lepton
pair with zero angular momentum. The matrix element of a vector
current coupling to the lepton pair vanishes and the decay is
dominated by $Z$-penguin and electroweak box contributions in the
SM. The axial-vector current coupling entails a $m_\ell^2/M_B^2$
helicity suppression which strongly suppresses the $B_q \to \ell^+
\ell^-$ branching ratio. The small size of the bottom-quark and lepton
Yukawa couplings renders the Higgs-penguin diagrams negligible in the
SM. This suppression mechanism does not hold for generic scalar
interactions and the contribution of scalar operators could
significantly increase the branching ratio above the SM expectation.
Together with the fact that the theory prediction of such branching
ratios is under good control, this enhancement makes $B_q \to \ell^+
\ell^-$ decays an ideal testing ground for new scalar interactions.

For the SM operators the branching ratio of $B_q \to \ell^+ \ell^-$,
$\mathcal{B}(B_q \to \ell^+ \ell^-)$, is given by
\begin{equation}
  \frac{G_F^2}{4 \pi} f_{B_q}^2
  M_{B_q}^5 \tau_{B_q} |V_{tb}^*V_{tq}|^2 \sqrt{\xi_{lq}}
  \left[\xi_{lq} C_S^2 + \left(C_P - \frac{2 m_\ell}{M_B^2} C_A
    \right)^2 \right] \, ,
\label{bll_brformula}
\end{equation}
where $\xi_{lq} = 1- 4 m_\ell^2/M_B^2$ is a kinematic
factor. Numerically, only the Wilson coefficient $C_A$ is
relevant. This coefficient has been computed at NLO in the
SM~\cite{Buchalla:1993bv,Misiak:1999yg,Buchalla:1998ba}. The
dependence on $f_{B_q}$ and the CKM parameters cancel in a combined
CKM fit, if the experimental information of $\Delta M_q$ is also taken
into account. The results of the CKMfitter group \cite{Charles:2004jd}
which were presented at ICHEP10 are
\begin{align}
\mathcal{B}(B_s \to \mu^+ \mu^-) &= 3.073^{+0.070}_{-0.190} \times
10^{-9} \\
\mathcal{B}(B_d \to \mu^+ \mu^-) &= 9.87^{+0.25}_{-0.67} \times 10^{-11} \; .
\end{align}

The $B_q \to \mu^+ \mu^-$ decays are sensitive to different extensions
of the SM: a modification of the SM $Z$-penguin can be measured using
the precise theory prediction of these modes, while the helicity
suppression provides an ideal test of flavour changing neutral scalar
interactions in the $C_S^{(\prime)}$ and $C_P^{(\prime)}$ Wilson
coefficients -- the extension of Eq.~(\ref{bll_brformula}) for
chirality-flipped operators has a similar structure and is given in
Ref.~\cite{Bobeth:2001jm}.

Potentially large flavour-changing scalar interactions appear in many
models with extended Higgs sectors. For example in the Minimal Flavour
Violation (MFV) variant of the MSSM the branching ratios could be
enhanced by several orders of magnitude. While the ratio
$\mathcal{B}(B_d \to \mu^+ \mu^-)/\mathcal{B}(B_s \to \mu^+ \mu^-)$
stays constant in models with MFV, this is no longer the case in more
generic extensions of the SM. The sensitivity of this ratio to generic
soft breaking terms in the MSSM can then be used to distinguish
different flavour models or falsify other extensions of the SM
\cite{Straub:2010ih}.

The strongest experimental upper bound on the $B_s$ decay,
$\mathcal{B}(B_s \to \mu^+ \mu^-) < 43 \times 10^{-9}$ at 95\% CL,
comes from the CDF collaboration using 3.7 fb$^{-1}$ of integrated
luminosity \cite{Abazov:2010fs}. The most recent result of the D0
collaboration gives an upper bound of $\mathcal{B}(B_s \to \mu^+
\mu^-) < 51 \times 10^{-9}$ at 95\% CL from 6.1 fb$^{-1}$ of data
\cite{CDFbsmm}. In the future ATLAS, CMS and LHCb will search for $B_s
\to \mu^+ \mu^-$ at the LHC. To measure such a small branching ratio,
good control of the background is needed ~\cite{Serra:2011ic}. The
main background source for all experiments is the combinatorial double
semi-leptonic decay $b \bar b \to \mu^+ \mu^- X$, where the two muons
accidentially form a secondary vertex.  Finally, at hadron colliders
only a relative measurement of the branching ratio is possible, where
a well known decay channel is used for normalization. If the decay
$B^+ \to J/\psi K^+$ is used as a normalization channel then the
uncertainty on the decay constant ratio $f_{B_d}/f_{B_s}$ introduces a
parametric uncertainty on the branching ratio of 15\%. Two strategies
to reduce this uncertainty are described in
Ref.~\cite{Serra:2011ic}. Using the modified frequentist approach the
90\% CL limits expected from the LHCb and CMS experiments for 1
fb$^{-1}$ integrated luminosity at a collision energy of 7 TeV are
$\mathcal{B}(B_s \to \mu^+ \mu^-) < 7.0 \times 10^{-9}$ and
$\mathcal{B}(B_s \to \mu^+ \mu^-) < 15.8 \times 10^{-9}$ respectively
\cite{Serra:2011ic}.

In the SM the branching ratio of the $D \to \ell^+ \ell^-$ decays is
controlled by long distance dynamics. The short distance $Z$-penguin
and box diagrams are CKM- and $m_b^2/M_W^2$-suppressed. To a good
approximation, the long distance dominated $D \to \gamma \gamma$ decay
gives the sole contribution to the branching ratio $\mathcal{B}(D \to
\mu^+ \mu^-) \sim 2.7 \times 10^{-5} \mathcal{B}(D \to \gamma \gamma)$
\cite{Burdman:2001tf}. To have a clear signal of new physics the
branching ratio has to lie significantly above the SM expectation
$\mathcal{B}(D \to \mu^+ \mu^-) \sim (2.7 - 8) \times 10^{-13}$.  Such
an enhancement is not easily achievable in typical extensions of the
SM -- see e.g. Ref.~\cite{Paul:2010pq}. The best experimental bounds
come from the Belle collaboration. Using 660 fb$^{-1}$ of data they
find $\mathcal{B}(D \to \mu^+ \mu^-) < 1.4 \times 10^{-7}$ and a
slightly better bound for the decay into an electron-positron
pair~\cite{Petric:2010yt} -- see also Ref.~\cite{Mohanty:2010ii} for
further information.

\section{Lepton flavour and number violating decays }

While lepton flavour and lepton number are good quantum numbers within
the SM (with massless neutrinos), lepton flavour is known to be
explicitly violated by neutrino mixing and there is no fundamental
symmetry principle which dictates charged lepton number or flavour
conservation in beyond-SM models.  Because the SM expectation for
these modes is vanishingly small, searches for lepton number violating
(LNV) or lepton flavour violating (LFV) decays are very sensitive
probes of physics beyond the SM, potentially at mass scales well
beyond the electroweak scale.  For most models the most sensitive
probes are in $\tau$ or $\mu$ decays, however LFV and LNV can also be
probed in leptonic and semileptonic decays of heavy flavour mesons,
providing a complementary approach to the $\tau$ searches, and in some
cases constraining specific scenarios beyond the reach of the $\tau$
modes.

Experimental searches for LFV and LNV decays in $B$ and $D$ meson
decays are performed either as incidental additions to related lepton
flavour and number conserving searches, or as dedicated searches for
specific modes.  The former is usually the case for LFV studies
involving first and second generation leptons, in particular $B^0 \to
e^\pm \mu^\mp$ and $B \to K^{(*)}/\pi e^\pm \mu^\mp$, while modes
involving $\tau$ leptons usual require dedicated searches.  The most
restrictive limits on these modes include $B(B \to K (K^*) e^\pm
\mu^\mp) < 5.1 (0.38) \times 10^{-7}$~\cite{babar_lfv} and $B(B^0 \to
e^{\pm} \mu^\mp) < 6.4 \times 10^{-8}$~\cite{cdf_lfv}.

Dedicated searches have been performed for modes with $\tau$ leptons:
$B^0 \to \tau^\pm \ell^\mp$ and $B \to K \tau^\pm \ell^\mp$ (where
$\ell = e, \mu$).  Missing energy associated with the $\tau$ decay
necessitates the use of hadronic tag reconstruction in these modes,
resulting in substantially reduced signal sensitivity compared with
the $e -\mu$ modes.  Current best limits on these modes are $B(B^0 \to
\tau^\pm e^\mp) < 2.8 \times 10^{-5}$, $B(B^0 \to \tau^\pm \mu^\mp) <
2.2 \times 10^{-5}$~\cite{babar_miika} and $B(B^+ \to K^+ \tau^\pm
\mu^\mp) < 7.7 \times 10^{-5}$ at $90\%$ confidence
level~\cite{babar_owen}.

Belle recently reported results of a search for the LFV and LNV modes
$B^+ \to D^- \ell^+ \ell '^+$ (where $\ell, \ell ' = e, \mu$),
motivated by a model with massive Majorana neutrinos~\cite{atre},
obtaining branching fraction limits ranging from $(1 - 3) \times
10^{-6}$~\cite{belle_LNV}.

\section{Rare Kaon decays}

The rare $K\to\pi\nu\bar\nu$, $K_L\to\pi^0\ell^+\ell^-$ and $K_L\to
\mu^+\mu^-$ decays play a central role in constraining and testing new
sources of flavour violation. In the SM, light-quark loops are
severely suppressed and loops with internal top quarks make the
dominant contributions to the branching ratios of these decays. New
physics would significantly alter the branching ratios, if the new
sources of flavour violation are not suppressed by a coupling constant
smaller than $V_{ts} V_{td}^* = \mathcal{O}(\lambda^5)$.

To exploit this new physics sensitivity a precise calculation of the
SM background is needed. The SM short distance contribution to the
branching ratio is above 90\% for the $K\to\pi\nu\bar\nu$ decays, but
below 50\% for the other modes. Correspondingly, the uncertainty in
the SM prediction for the $K_L\to\pi^0\ell^+\ell^-$ and $K_L\to
\mu^+\mu^-$ decays is dominated by the long distance calculations.

The rare decay $K_L\to \mu^+\mu^-$ was important in establishing the
GIM mechanism. No single photon penguin contributes to the decay.
Accordingly, at leading order in the electroweak couplings, only
$Z$-penguin and box diagrams can contribute. In these diagrams light
quark dynamics are suppressed by a power-like GIM mechanism. At the
next order in $\alpha$ the two-photon penguin results in a long
distance contribution, where the absorptive part saturates the
measured branching ratio and the dispersive part can only be estimated
with large uncertainties~\cite{Isidori:2003ts}. Even with these large
theoretical uncertainties, important constraints on flavour changing
scalar operators result~\cite{Mescia:2006jd} from the measurement
$\mathcal{B}(K_L\to \mu^+\mu^-) = 6.84(11) \times
10^{-9}$~\cite{Nakamura:2010zzi}.

The rare $K_L \to \pi^0 \ell^+\ell^-$ decays are CP violating at
leading order in the electroweak interactions. The direct CP violating
contribution is dominated by top-quark loops in the SM and is
correspondingly sensitive to short distances. There is a long distance
pollution via indirect CP violation through $K_L$--$K_S$ mixing and
the CP conserving two-photon contribution. Contrary to the $K_L\to
\mu^+\mu^-$ decay, these long distance effects can be calculated in
chiral perturbation theory. A combined measurement could be used to
disentangle short- and long-distance contributions and would be
sensitive to the scalar operators of Eq.~(\ref{eq:ops}) or tensor
operators~\cite{Mescia:2006jd}.

Rare Kaon decays with neutrinos in the final state are exceptionally
clean. Only $Z$-penguin and box diagrams contribute and light-quark
effects are suppressed by a quadratic GIM mechanism. The branching
ratio of the $K^+\to\pi^+\nu\bar{\nu}$ decay is given,
\begin{equation}
  \label{eq:brkplus}
  \kappa_+ (1+\Delta_{\text{EM}})
  \Bigg[\left(\frac{\text{Im}\lambda_t}{\lambda^5} X_t\right)^2 +
  \left(\frac{\text{Re}\lambda_c}{\lambda} \left(P_c + \delta P_{c,u}
    \right) + \frac{\text{Re}\lambda_t}{\lambda^5} X_t\right)^2
  \Bigg] \, ,
\end{equation}
where the normalisation factor $\kappa_+$ is extracted from $K_{l3}$
decays including isospin breaking corrections and the corresponding
photon cutoff dependence denoted by
$\Delta_{\text{EM}}$~\cite{Mescia:2007kn}. The factors $P_c$ and
$\delta P_{c,u}$ comprise the short- and long-distance effects of
light quarks -- for the most recent theoretical estimates see
Refs.~\cite{Brod:2008ss,Isidori:2005xm}.  The Wilson coefficient $X_t$
is the top-quark contribution of $C_L^{sd}$ times $2 \pi \sin^2
\theta_W/\alpha$. It is now known to high accuracy, after the
inclusion of two-loop electroweak corrections removed the electroweak
scheme dependence of the electroweak input parameters
\cite{Brod:2010hi,Stamou:2011aj}. This error reduction is even more
important for the CP violating neutral decay mode whose branching
ratio depends only on the top-quark contribution. The errors in the SM
theory prediction $\mathcal{B}(K^+\to\pi^+\nu\bar\nu) =
(7.81^{+0.80}_{-0.71}\pm 0.29) \times 10^{-11}$ and
$\mathcal{B}(K_L\to\pi^0\nu\bar\nu) = (2.43^{+0.40}_{-0.37}\pm 0.06)
\times 10^{-11}$ are dominated by the parametric uncertainties (first
error), while the intrinsic theory error (second error) is less than
4\% for both modes.

The precise theory prediction and the $V_{ts}^* V_{td}$ suppression
imply a high sensitivity to new physics with generic flavour
violation. Effects of new heavy particles modify $X_t = 2 \pi \sin^2
\theta_W/\alpha(C_L^{sd} + C_R^{sd})$, and correspondingly the
branching ratios of the neutral and charged decay modes. Correlations
between the two branching ratios can then be used to discriminate
between various new physics models e.g. Randal-Sundrum type models,
theories with extra fermions or the
MSSM~\cite{Straub:2010ih}. Interactions of additional light particles
could also be tested via the $K \to \pi + \textrm{nothing}$ signals --
for details see Ref.~\cite{Smith:2010st}.

The NA62 experiment aims to measure 80 $K^+\to\pi^+\nu\bar{\nu}$
events with signal acceptance and background both of the order of
10\%~\cite{Spadaro:2011ue}, using an 75GeV, 800MHz beam with $\sim
6\%$ $K^+$. Several background rejection methods have to be employed
to measure a branching ratio of $\mathcal{O}(10^{-10})$: a cut on the
missing mass of the reconstructed candidates can be performed by
measuring the momentum of the incoming $K^+$ momentum with the
so-called Gigatracker, and the momenta of the daughter particles with
a straw-chamber magnetic spectrometer. Further background from
leptonic and semileptonic Kaon decays is removed using a Cherenkov
detector, while a photon-veto system should remove decay modes with
$\pi^0$s and/or radiative photons.

\section{Conclusions}

Recent progress from experiment and theory in the field of rare decays
has confirmed that flavour and CP violation is well described by the
CKM mechanism. However, at present, only a limited number of
theoretically clean observables have been probed by
experiments. Current and future experiments at CERN, KEK and SuperB
will change this and allow precision tests of flavour violation even
beyond the Standard Model. Rare decays will therefore continue to be
an important tool for understanding fundamental interactions.

\section*{Acknowledgements}
We would like to thank the organisers of CKM2010 for the interesting
time in Warwick. MG would like to thank Joachim Brod for comments on
parts of the manuscript.


\begin{thebibliography}{30}

\bibitem{Glashow:1970gm}
  S.~L.~Glashow, J.~Iliopoulos and L.~Maiani,
  Phys.\ Rev.\  D {\bf 2} (1970) 1285.

\bibitem{Kobayashi:1973fv}
  M.~Kobayashi, T.~Maskawa,
  Prog.\ Theor.\ Phys.\  {\bf 49 } (1973)  652-657.


\bibitem{Misiak:2006zs}
  M.~Misiak {\it et al.},
  Phys.\ Rev.\ Lett.\  {\bf 98} (2007) 022002.

\bibitem{Bobeth:2003at}
  C.~Bobeth, P.~Gambino, M.~Gorbahn and U.~Haisch,
  JHEP {\bf 0404} (2004) 071.

\bibitem{Paz:2010wu}
  G.~Paz,
  [arXiv:1011.4953 [hep-ph]].

\bibitem{Benzke:2010tq}
  M.~Benzke, S.~J.~Lee, M.~Neubert and G.~Paz,
  [arXiv:1012.3167 [hep-ph]].


\bibitem{Beneke:2004dp}
  M.~Beneke, T.~Feldmann,  D.~Seidel,
  Eur.\ Phys.\ J.\  C {\bf 41} (2005) 173.
  [arXiv:hep-ph/0412400].

\bibitem{Khodjamirian:2011gy}
  A.~Khodjamirian,
  [arXiv:1101.2328 [hep-ph]].

\bibitem{Liu:2011ra}
  Z.~Liu, S.~Meinel, A.~Hart {\it et al.},
  [arXiv:1101.2726 [hep-ph]].

\bibitem{Khodjamirian:2010vf}
  A.~Khodjamirian, T.~Mannel, A.~A.~Pivovarov and Y.~M.~Wang,
  JHEP {\bf 1009} (2010) 089.

\bibitem{Egede:2010zc}
  U.~Egede, T.~Hurth, J.~Matias {\it et al.},
  JHEP {\bf 1010} (2010) 056; 
  U.~Egede, T.~Hurth, J.~Matias {\it et al.},
  [arXiv:1012.4603 [hep-ph]].


\bibitem{babar} B. Aubert {\it et~al.} (\babar\ collaboration), Phys.\ Rev.\ {\bf  D79},  031102 (2009).
\bibitem{belle} J.T. Wei {\it et~al.} (Belle collaboration), Phys.\ Rev.\ Lett.\ {\bf 103}, 171801 (2009).
\bibitem{cdf} T. Aaltonen  {\it et~al.} (CDF collaboration), CDF note 10047 (2010).

\bibitem{Eigen:2011rq}
  G.~Eigen,
  [arXiv:1101.0470 [hep-ex]].

\bibitem{Burdman:1998mk}
  G.~Burdman,
  Phys.\ Rev.\  D {\bf 57} (1998) 4254.

\bibitem{Beneke:2001at}
  M.~Beneke, T.~Feldmann,  D.~Seidel,
  Nucl.\ Phys.\  B {\bf 612} (2001) 25.

\bibitem{Alok:2009tz}
  A.~K.~Alok {\it et~al.},
  JHEP {\bf 1002} (2010) 053.

\bibitem{Feldmann:2002iw}
  T.~Feldmann and J.~Matias,
  JHEP {\bf 0301} (2003) 074.


\bibitem{Skottowe:2010bs}
  H.~Skottowe, 
  [arXiv:1005.2433 [hep-ex]].

\bibitem{Soomro:2011vn}
  F.~Soomro,
  [arXiv:1101.5717 [hep-ex]].

\bibitem{roadmap} B.~Adeva {\it et~al.} (LHCb collaboration), Roadmap for selected key measurements of LHCb, arXiv 0912.4179.

\bibitem{belle_incl_bsgamma} A.\ Limosani {\it et~al.} (Belle collaboration), 
Phys.\ Rev.\ Lett.\  {\bf 103} 241801 (2009).

\bibitem{wenfang}
  W.~Wang,
  [arXiv:1102.1925 [hep-ex]].

\bibitem{belle_excl_bdgamma} D.\ Mohapatra {\it et~al.} (Belle collaboration), 
Phys.\ Rev.\ Lett.\  {\bf 96}, 221601 (2006).

\bibitem{babar_excl_bdgamma} B.\ Aubert {\it et~al.} (\babar collaboration), 
Phys.\ Rev.\ Lett.\ {\bf 98}, 151802 (2007).

\bibitem{babar_debbie} P.\ del Amo Sanchez {\it et~al.} (\babar collaboration), 
Phys.\ Rev.\  {\bf D 82}, 051101 (2010).

\bibitem{ball_bdgamma} P.\ Ball, G.\ Jones and R.\  Zwicky, Phys.\ Rev.\ {\bf D 75}, 054004 (2007).

\bibitem{buchalla} G. Buchalla {\it et~al.}, Phys.\ Rev.\ {\bf D63}, 014015 (2001). 
\bibitem{hou} A. Hovhannisyan, W. S. Hou and N. Mahajan, Phys.\ Rev.\ {\bf D 77}, 014016 (2008).
\bibitem{kruger2} F. Kr\"uger and J. Matias, Phys.\ Rev.\ {\bf D71}, 094009 (2005).

\bibitem{Kamenik:2010na}
  J.~F.~Kamenik,
  [arXiv:1012.5309 [hep-ph]].

\bibitem{Smith:2010st}
  C.~Smith,
  [arXiv:1012.4398 [hep-ph]].

\bibitem{babar_SL_Knunu} 
  P. del Amo Sanchez {\it et~al.} (\babar\ collaboration), Phys.\ Rev.\ {\bf D82}, 112002 (2010); 
  
\bibitem{exp_knunu}
   B. Aubert {\it et~al.}  (\babar\ collaboration), Phys.\ Rev.\ {\bf D78}, 072007, (2008); 
    K. F. Chen {\it et~al.}  (BELLE collaboration), Phys.\ Rev.\ Lett.\ 99, 221802 (2007); 
    B. Aubert {\it et~al.}  (\babar collaboration), Phys.\ Rev.\ Lett.\ 94, 101801 (2005).



\bibitem{Buchalla:1993bv}
  G.~Buchalla and A.~J.~Buras,
  Nucl.\ Phys.\  B {\bf 400} (1993) 225.

\bibitem{Misiak:1999yg}
  M.~Misiak and J.~Urban,
  Phys.\ Lett.\  B {\bf 451} (1999) 161.

\bibitem{Buchalla:1998ba}
  G.~Buchalla and A.~J.~Buras,
  Nucl.\ Phys.\  B {\bf 548} (1999) 309.

\bibitem{Charles:2004jd}
  J.~Charles {\it et al.}  [CKMfitter Group],
  Eur.\ Phys.\ J.\  C {\bf 41} (2005) 1
  updated results and plots available at: http://ckmfitter.in2p3.fr.

\bibitem{Bobeth:2001jm}
  C.~Bobeth, A.~J.~Buras, F.~Kruger and J.~Urban,
  Nucl.\ Phys.\  B {\bf 630} (2002) 87.

\bibitem{Straub:2010ih}
  D.~M.~Straub,
  arXiv:1012.3893 [hep-ph].

\bibitem{Abazov:2010fs}
  V.~M.~Abazov {\it et al.} [ D0 Collaboration ],
  Phys.\ Lett.\  {\bf B693 } (2010)  539-544.


\bibitem{CDFbsmm}
  CDF Collaboration,
  CDF Public Note 9892.

\bibitem{Serra:2011ic}
  N.~Serra,
  [arXiv:1102.2410 [hep-ex]].


\bibitem{Burdman:2001tf}
  G.~Burdman, E.~Golowich, J.~L.~Hewett and S.~Pakvasa,
  Phys.\ Rev.\  D {\bf 66} (2002) 014009.

\bibitem{Paul:2010pq}
  A.~Paul, I.~I.~Bigi and S.~Recksiegel,
  Phys.\ Rev.\  D {\bf 82} (2010) 094006
  [Erratum-ibid.\  D {\bf 83} (2011) 019901]. 

\bibitem{Petric:2010yt}
  M.~Petric {\it et al.}  [Belle Collaboration],
  Phys.\ Rev.\  D {\bf 81} (2010) 091102.

\bibitem{Mohanty:2010ii}
  G.~B.~Mohanty,
  [arXiv:1012.1930 [hep-ex]].



\bibitem{babar_lfv} B. Aubert {\it et~al.} (\babar\ collaboration), Phys.\ Rev.\ {\bf D73}, 092001 (2006).

\bibitem{cdf_lfv} T. Aaltonen {\it et~al.} (CDF collaboration), Phys.\ Rev.\ Lett.\ 102, 201801 (2009).

\bibitem{babar_miika} 
  B. Aubert {\it et~al.} (\babar\ collaboration), Phys.\ Rev.\ {\bf D77}, 091104, (2008).

\bibitem{babar_owen}
  B. Aubert {\it et~al.}  (\babar\ collaboration), Phys.\ Rev.\ Lett.\ 99, 201801, (2007).
  
\bibitem{atre}
  A. Atre {\it et~al.} , JHEP 0905, 030 (2009); J.M. Zhang and G.L. Wang,
  G. Cvetic {\it et~al.} , Phys.\ Rev.\ {\bf D82}, 053010 (2010).

\bibitem{belle_LNV}
  K. Hayasaka, talk on behalf of the Belle Collaboration at ICHEP2010,
  Paris.



\bibitem{Isidori:2003ts}
  G.~Isidori and R.~Unterdorfer,
  JHEP {\bf 0401} (2004) 009.


\bibitem{Mescia:2006jd}
  F.~Mescia, C.~Smith, S.~Trine,
  JHEP {\bf 0608 } (2006)  088.


\bibitem{Nakamura:2010zzi}
  K. Nakamura {\it et al.} [ Particle Data Group Collaboration ],
  J.\ Phys.\ G {\bf G37 } (2010)  075021.


\bibitem{Mescia:2007kn}
  F.~Mescia and C.~Smith,
  Phys.\ Rev.\  D {\bf 76} (2007) 034017.

\bibitem{Brod:2008ss}
  J.~Brod and M.~Gorbahn,
  Phys.\ Rev.\  D {\bf 78} (2008) 034006.

\bibitem{Isidori:2005xm}
  G.~Isidori, F.~Mescia and C.~Smith,
  Nucl.\ Phys.\  B {\bf 718} (2005) 319.

\bibitem{Brod:2010hi}
  J.~Brod, M.~Gorbahn and E.~Stamou,
  Phys.\ Rev.\  D {\bf 83} (2011) 034030.  

\bibitem{Stamou:2011aj}
  E.~Stamou,
  [arXiv:1101.3245 [hep-ph]].

\bibitem{Spadaro:2011ue}
  T.~Spadaro,
  [arXiv:1101.5631 [hep-ex]].

\end{thebibliography}
\end{document}